\documentclass[12pt,preprint]{aastex}
\usepackage{epsfig}
\usepackage{natbib}

\newcommand{\kms}{km~s$^{-1}$}

\newcommand{\ergsa}{ergs~cm$^{-2}$~s$^{-1}$~\AA$^{-1}$}

\newcommand{\hst}{\textit{HST\/}}

\def\deg{\hbox{$^\circ$}}

\shorttitle{}
\shortauthors{Feldman et al.}

\begin{document}

\title{Hubble Space Telescope Observations of Comet 9P/Tempel 1 during
the Deep Impact Encounter}  

\author{Paul D. Feldman, Stephan R. McCandliss, Matthew Route}

\affil{Department of Physics and Astronomy, The Johns Hopkins University\\ 
Charles and 34th Streets, Baltimore, MD 21218-2695}
\email{pdf@pha.jhu.edu}

\author{Harold A. Weaver}

\affil{Space Department,
Johns Hopkins University Applied Physics \mbox{Laboratory,}\\
11100 Johns Hopkins Road, Laurel, MD 20723-6099}

\author{Michael F. A'Hearn}

\affil{Department of Astronomy, University of Maryland, \\
College Park MD 20742-2421}

\author{Michael J. S. Belton}

\affil{Belton Space Exploration Initiatives,
Tucson, AZ 85716}

\author{Karen J. Meech}

\affil{University of Hawaii, Institute for Astronomy and \\
NASA Astrobiology Institute, \\
2680 Woodlawn Drive, Honolulu, HI 96822}



\clearpage
\begin{abstract}

We report on the {\it Hubble Space Telescope} program to observe
periodic comet 9P/Tempel 1 in conjunction with NASA's Deep Impact
mission. Our objectives were to study the generation and evolution of
the coma resulting from the impact and to obtain wide-band images of
the visual outburst generated by the impact.  Two observing campaigns
utilizing a total of 17 \hst\ orbits were carried out: the first
occurred on 2005 June 13--14 and fortuitously recorded the appearance
of a new, short-lived fan in the sunward direction on June 14. The
principal campaign began two days before impact and was followed by
contiguous orbits through impact plus several hours and then snapshots
one, seven, and twelve days later.  All of the observations were made
using the Advanced Camera for Surveys (ACS).  For imaging, the ACS High
Resolution Channel (HRC) provides a spatial resolution of 36 km (16 km
pixel$^{-1}$) at the comet at the time of impact.  Baseline images of
the comet, made prior to impact, photometrically resolved the comet's
nucleus.  The derived diameter, 6.1 km, is in excellent agreement with
the $6.0 \pm 0.2$~km diameter derived from the spacecraft imagers.
Following the impact, the HRC images illustrate the temporal and
spatial evolution of the ejecta cloud and allow for a determination of
its expansion velocity distribution.  One day after impact the ejecta
cloud had passed out of the field-of-view of the HRC.

\end{abstract}


\newpage
\section{INTRODUCTION}

The Deep Impact mission \citep{A'Hearn:2005a,A'Hearn:2005b}
successfully placed a 364 kg, about half copper, impactor, onto the
surface of comet 9/P Tempel 1 at a relative velocity of 10.3~\kms\ on
2005 July 4 at 05:52:03 UT (as seen from Earth).  The event was
observed by cameras aboard the mother spacecraft and by a large number
of Earth- and space-based telescopes as part of an extensive campaign
to study the comet prior to, during, and in the course of several days
following the impact \citep{Meech:2005}.  The {\it Hubble Space
Telescope} (\hst) provided the highest spatial resolution images from
Earth (36 km, 16 km pixel$^{-1}$) using the Advanced Camera for Surveys
(ACS) High Resolution Channel (HRC), and also ultraviolet imaging using
the Solar Blind Channel (SBC) of ACS to study the evolution of gaseous
species, particularly CO, released by the impact.  This paper describes
the \hst\ campaign and presents the images obtained and the preliminary
interpretation of the data in terms of the physical properties of the
observed ejecta.  We also compare the ejecta produced by the impact
with that from a ``natural outburst'' that was serendipitously observed
on 2005 June 14.

\section{OBSERVATIONS}

The \hst\ campaign consisted of 17 separate ``visits'', each comprising
a single \hst\ orbit allowing $\sim$53~minutes of target visibility per
orbit.  Twelve of these orbits were dedicated to visible
imaging with the HRC and they are summarized in Table~\ref{obs_table}.
Four of the visits (program ID 10144) were on 2005 June 13--14, three
weeks before impact, and spanned a full rotation (40.7 h) of the
nucleus.  The remainder (program ID 10456) were grouped into three
periods: a pre-impact group beginning roughly one cometary rotation
before impact to establish a baseline for the data to follow; an orbit
that included the impact time; and several orbits immediately following
the impact and continuing, with single orbits, 7 and 12 days after the
impact.  The importance of the \hst\ observations was underscored by
the coordination between the STScI and JPL to adjust the impact time to
allow for \hst\ visibility of the event.  As it turned out, impact
occured 15 minutes before the comet set below {\it HST's} horizon.

\placetable{obs_table}

All of the visible imaging was done with the HRC using a broad V
filter, F606W, to maximize the sensitivity to faint coma structures.
This filter is centered at a wavelength of 5907~\AA\ and has a bandpass
of 2342~\AA.\footnote[1]{These parameters are from the ACS 
Instrument Handbook available at \\ 
http://www.stsci.edu/hst/acs/documents/handbooks/cycle15/c05\_imaging2.html\#357257.  
For purposes of photometric calibration, for the F606W filter used with
the HRC, \citet{Sirianni:2005} define a pivot wavelength of
5888~\AA\ and a width of 665~\AA, where the latter includes the system
throughput as a multiplicative factor.} The observing program, with the
exception of the impact orbit, consisted of pairs of long (140 or
300~s) and short (40~s) exposures interlaced with objective prism
images (PR200L) that have so far been difficult to interpret.  For the
impact orbit, a continuous sequence of 40 second exposures using the
HRC $512 \times 512$ pixel mode (half-frame; the full-frame $1024
\times 1024$ pixel mode was used for all other July observations) was
used to avoid the possibility of filling the data buffer and requiring
a buffer dump at the time of impact.  With overhead included, the
images were taken at 75-second intervals.  The impact occurred almost
exactly in the middle of exposure J9A805EBQ.  A total of 12 exposures
were taken including and following impact.

Because of the large overhead in switching cameras during a single
\hst\ orbit, entire visits were dedicated to the SBC observations.  The
results of these observations are being reported separately 
\citep{Feldman:2006}.

\section{VISUAL IMAGING}

All of the individual images were processed with pipeline software to
produce flat-fielded, geometrically corrected output products that were
then rotated to orient celestial north up.  An example of the nature of
the appearance of the ejecta plume is shown in Fig.~\ref{full} in which
the right-hand image, a composite of two 140-s exposures, was taken
about one hour after impact while the left-hand image, a composite of
four 140-s exposures, was taken 54 hours earlier.  The pre-impact image
is dominated by light from the nucleus and the inner coma (because the
brightness of the quiescent coma falls as the inverse of the distance
from the nucleus) while the post-impact image shows clearly the
extended fan of ejected material.  To enhance the effects of the
impact, the remainder of the images presented will be the ratio of a
given image to a pre-impact image.  The false color scale will then
correspond to the relative increase of material in the coma as a result
of the impact relative to its quiescent state.

\placefigure{full}

\subsection{Pre-impact Images}

\subsubsection{Radial Profiles and the Photometric Resolution of the
Nucleus}

The HRC images of the quiescent comet are dominated by a strong central
peak, surrounded by the dust coma, which while not perfectly
azimuthally symmetric, nevertheless closely follows a $\rho^{-1}$
brightness distribution as a function of distance from the nucleus.
The width of the bright peak closely matches the camera's point spread
function (PSF) (FWHM of 2.25 pixels or $\sim$36~km projected at the
comet for the July observations).  Most of the signal is from the
nucleus, which is not spatially resolved.  For comparison with the
images from the Deep Impact spacecraft, we used the last HRC image
prior to impact, J9A805EAQ, to evaluate the photometric area of the
nucleus.  An azimuthally symmetric coma model, convolved with the
instrument PSF, was subtracted from the image and the azimuthally
averaged radial profile extracted, as has been done for many comets
observed by \hst\ \citep{Lamy:2001,Lamy:2004}.  The result is shown in
Fig.~\ref{nucleus}, where the nucleus is found to account for
$\sim$90\% of the counts in the central pixel.  Using the recent ACS
calibration data of \citet{Sirianni:2005}, the V magnitude of the
nucleus is 17.49.  Assuming an average geometric albedo of 4.0\% and a
phase law of 0.04 mag deg$^{-1}$ \citep{Lamy:2004}, this magnitude
implies an effective diameter of 6.1~km.  Direct imaging of the nucleus
from Deep Impact gives a diameter of $6.0 \pm 0.2$~km
\citep{A'Hearn:2005b}, which suggests that our choices for the albedo
and phase law are accurate.

\placefigure{nucleus}

\subsubsection{The June 14 Outburst}

Our intention for the June visits was to obtain HRC images evenly
spaced over a single rotation period, 40.7 h \citep{A'Hearn:2005b}, to
compare with images taken by the spacecraft cameras that would yield
comparable spatial resolution at the distance of the comet.  It was
also an opportunity to test our strategy of taking half-frame images
during the impact orbit and verify the ephemeris.  The four visits are
summarized in Table~\ref{obs_table}.  The fourth visit shows a markedly
different coma shape from that seen in the preceding visits, and the
ratio of these images to one from the third visit, 7 hours earlier, is
shown in Fig.~\ref{june}.  The ratio images were created from sums
of two 300-second exposures.  Fortuitously, the fourth visit was split
over two \hst\ orbits so that the two images shown in the figure were
taken 67 minutes apart.  The expansion of the outburst cone, filling
nearly the entire sunward hemisphere, is clearly seen by comparison of
the two images, shown as contour plots in Fig.~\ref{contour_june}.
The outer isophote in each figure is well matched by a semi-circle
centered at the nucleus, suggesting uniform radial expansion into the
sunward hemisphere.  The velocity of this isophote, projected onto the
plane of the sky, is 145~m~s$^{-1}$, which for a hemispherical shell
represents a true expansion velocity.  As this isophote contains 98.5\%
of the additional light produced by the outburst, the derived velocity
is a good approximation to the maximum expansion velocity of the
outflowing material.  For a hemispherical shell, the peak brightness
occurs at the inner boundary of the shell, which leads to a minimum
velocity of 60~m~s$^{-1}$.  The similar morphology of the contours in
Fig.~\ref{contour_june} suggests that the shell is expanding uniformly
and that the event that produced the shell was short-lived relative to
the time since the outburst.

\placefigure{june}
\placefigure{contour_june}

Extrapolating backwards to the nucleus, assuming constant velocity, the
outburst occurred $\sim$4.5~hours before the mid-point of the left
exposure of Fig.~\ref{june}, or at $\sim$09:45~UT.  This time appears
to be consistent with the same event seen by the imagers on Deep
Impact, following a re-analysis of the data presented by
\citet{A'Hearn:2005b} and is also the same event that was reported by
\citet{Lara:2006}, who observed it at $\sim$21:00 UT on June 14.  A
duration time of the event, $\sim$10~minutes, deduced for several
natural outbursts from the Deep Impact light curve \citep[Fig.~4 of][]
{A'Hearn:2005b}, is consistent with the analysis presented above.

\subsection{Impact Orbit}

As noted above, for the impact orbit, a sequence of 40-second exposures
using the HRC $512 \times 512$ pixel mode was taken at 75~s intervals.
The impact occurred almost exactly in the middle of exposure
J9A805EBQ.  A total of 12 exposures were taken including and following
impact and these are shown, as ratios to the mean of the last four
exposures prior to impact in Fig.~\ref{visit05}.  The expansion of the
ejecta is illustrated by contour plots of four of these ratio images
given in Fig.~\ref{contour05}.  These plots illustrate the initial
expansion of the ejecta, characterized by a steady increase in
brightness in the vicinity of the nucleus, together with an apparent
elongation and displacement towards the southeast.  This is in a
direction opposite from a position angle between that of the projection of
the impactor velocity vector on the sky (355\deg) and that of its
downrange component tangent to the comet's surface (294\deg).  These
directions, as seen from Earth, are given by \citet{Carcich:2006} from
a recent shape model derived from the {\it in situ} imaging.  The
brightening in the uprange direction may be the later evolution of a
highly foreshortened uprange plume seen in the first second after
impact by the Deep Impact Medium Resolution Imager
\citep{A'Hearn:2005b}.  These authors note that such a plume is similar
to plumes observed in the laboratory in which the target material is
porous \citep{Schultz:2005}.

\placefigure{visit05}
\placefigure{contour05}

The outermost contour in the SE quadrant of the final image of the
sequence (J9A805EMQ) corresponds to an expansion velocity of
390~m~s$^{-1}$, assuming that the observed material was generated at
impact time.  Taking the difference of this image with a pre-impact
image, we find that 10\% of the additional light produced by the impact
corresponds to material with velocities higher than this, and 1\%
corresponds to velocities $\geq 1.5$~\kms.  This last component is
likely due to material produced in the initial ejecta plume
\citep{A'Hearn:2005b}.  Further discussion of these images, based on
the light curves derived from aperture photometry, is given below.

\subsection{Post-impact Images and Dust Expansion Velocity}

On all of the \hst\ orbits dedicated to HRC imaging post-impact, pairs
of images were taken with exposure times of 40 and 140 seconds each to
insure against possible saturation of the image due to a bright ejecta
plume near the nucleus.  The quiescent level of the comet was
established from the June observations.  These were interspersed with
prism images, so that each orbit can be represented by a pair of
images, each of 280~s exposure time, separated by 24 minutes.  In this
way, the temporal development of the ejecta morphology may be
illustrated, as in Fig.~\ref{visit06}, which shows the ratio images
from visit~06 beginning about one hour after impact.  Most striking in
this pair of images is the expansion of the ejecta ``fan'' in the SW
quadrant, in contrast to the images obtained during the 15 minute
period following the impact.  The fan is symmetric about a position
angle of 240\deg, which is close to the direction of the normal to the
surface of the comet at the impact site as seen from Earth (position
angle 233\deg) according to the comet shape model of
\citet{Carcich:2006}.  This fan geometry is characteristic of the
larger scale ground-based images that were obtained during the Deep
Impact campaign \citep{Schleicher:2006,Sugita:2005}.

\placefigure{visit06}

A similar pair, from an orbit 3.2~h later (visit~08), is shown in
Fig.~\ref{visit08}.  Here the morphology is almost identical except
that the scale of the images is about four times larger.  This implies
uniform expansion of the ejecta with a distribution of velocities that
remains constant with time.  As in the case of the June 14 outburst,
the maximum expansion velocity, associated with the smallest grains, can
be determined from the differences in the outermost contour in the
two panels derived from Fig.~\ref{visit06}, shown in
Fig.~\ref{contour06}.  The contours are well matched by semi-circular
isophotes, except for the NW quadrant where the effects of solar
radiation pressure are beginning to be felt \citep{Schleicher:2006}.
The outermost isophote contains 97\% of the light produced by the
impact in the SW quadrant.  The velocity derived from this isophote is
280~m~s$^{-1}$ and this value is consistent with the extrapolation to
both the impact time and to the outer contours of the images in
Fig.~\ref{visit08}.  For an expanding hemispherical shell, the inner
boundary of the shell, corresponding to the slowest moving particles,
is at the position of the brightest contour.  This corresponds to a
velocity of $\sim$80~m~s$^{-1}$.  In the images taken immediately
following impact (Fig.~\ref{contour05}), the contours in the SW
quadrant also appear to be expanding with a velocity of
$\sim$300~m~s$^{-1}$, suggesting that any acceleration of the excavated
material must take place within 100~km of the nucleus.

\placefigure{visit08}
\placefigure{contour06}

If we assume that the time during which material is ejected is short
compared to some later time after the event, then the radial brightness
profile will reflect the velocity distribution of the ejected
material.  In Fig.~\ref{rad06} we show radial profiles of both the pre-
and post-impact images used in the ratio images of Fig.~\ref{visit06}.
These profiles are generated by integrating the detector counts along
annuli between position angles $\pm 60$\deg\ from the symmetry axis.
The difference between these profiles, multiplied by the radius (in
pixel units), is shown in blue.  The dashed curve indicates what
fraction of the flux lies outside a given radial distance.

\placefigure{rad06}

The blue curves in Fig.~\ref{rad06} are converted to velocity by
dividing by the time difference between the exposure mid-point and the
impact time.  They are then normalized to unity with respect to
integration over velocity, and are shown in Fig.~\ref{velocity}.  The
two distribution functions, based on images with exposure mid-points
68.2 and 92.2 minutes, respectively, after the impact, are in good
agreement, the slight differences likely due to the finite time for
particle ejection (estimated to be 4--5~minutes), and the fact that the
integration over velocity does not fully extend to infinity.  With
these distribution functions nearly identical results are obtained for
the mean expansion speed, 115~m~s$^{-1}$, and rms velocity,
145~m~s$^{-1}$.  The most probable velocity is 70--80~m~s$^{-1}$, as
found from the contour plots.  The orbit 3.2~h later
(Fig.~\ref{visit08}) gives a qualitatively similar result but is
affected by the finite size of the detector.

\placefigure{velocity}

Our results may be compared with a number of other estimates of dust
expansion velocity.  From visible imaging, \citet{Schleicher:2006} report
a maximum velocity of 230~m~s$^{-1}$, while
\citet{Kuppers:2005}, from images taken with the OSIRIS camera on {\it
Rosetta}, report a typical velocity of 110~m~s$^{-1}$ with a maximum of
300~m~s$^{-1}$.  \citet{Sugita:2005}, from 8--13~$\mu$m images, derive
a mean expansion velocity of $125 \pm 10$~m~s$^{-1}$, which is consistent
with the larger particles seen in the infrared.  We also note that our
derived velocity is about twice as large as that seen in the June 14
outburst, an event presumably driven by the sublimation of volatile
ices, so that the energy for the plume expansion must have come mainly
from the impact.

The angular distribution of the ejecta material about the normal to
the surface of the comet at the impact site provides information
about the nature of the cratering process.  A significant change
in morphology is seen between the images taken during the first
14~minutes after impact (Fig.~\ref{visit05}) and those beginning
an hour after impact (Fig.~\ref{visit06}).  An example of the azimuthal
distribution of the ejecta in this latter case is shown in
Fig.~\ref{az06}.

\placefigure{az06}

By the time of the next visit with HRC exposures, $\sim$19~h after
impact, the ejecta plume had expanded beyond the $29'' \times 26''$
field-of-view of the camera.  The brightness surrounding the nucleus
remains higher than in the pre-impact images because the line-of-sight
passes through the ejecta cone in front of and behind the plane of the
sky.

\subsection{Light Curves}

Aperture photometry was generated from the images by integrating the
detector count rate over a range of circular apertures with radii from
2.5~pixels (40~km) to 250~pixels (4050~km) centered on the comet's
nucleus.  The data from the impact orbit (visit 05) are shown in the
left-hand panel of Fig.~\ref{impact_lc}, in which an average of
pre-impact points has been subtracted from the count rate integrated
over each aperture.  The impact flash may be responsible for the small
signal rise in the image containing the impact event.  This is followed
by a sharp increase in all of the apertures which then continue to
increase monotonically with time.  The increase in the three smallest
apertures is a factor of 8--10 over the pre-impact count rate in a
13.7~minute period.  The upward curvature seen in the larger apertures
may be the result of the fragmentation of the ejected grains into
smaller particles.  The near-linear rise in the 40~km radius aperture
is likely due to decreasing optical depth in the central region of the
aperture so that the effective scattering surface is increasing as the
material approaches optical depth unity.  There is no evidence in these
data for a dip in brightness increase at $\sim$200~s, as reported by
\citet{Keller:2005} from OSIRIS images, but this discrepancy may be the
result of the different viewing geometry from Earth and from the {\it
Rosetta} spacecraft.

\placefigure{impact_lc}

The right-hand panel of the figure shows that activity had ceased by
60~minutes after impact and that the ejecta had effectively moved out
of the fields-of-view of the three smallest apertures.  Only the
400~km radius aperture shows an increase from the last data point in
the left-hand panel, implying that the velocity of the slowest moving
ejecta particles was $<$135~m~s$^{-1}$.  From the 160~km aperture, we
also find this velocity must be $>$55~m~s$^{-1}$, consistent with the
value found from the contour analysis above.  The 49~minute data gap
makes it difficult to further constrain these values.

A lightcurve through 20~hours after impact, giving the observed count
rates in the larger apertures to facilitate comparison with ground-based
data, is shown in Fig.~\ref{post_lc}.  By the time of the following
\hst\ visit, 7.5~days after impact, the comet had returned to its
pre-impact level of activity and the photometry showed a slow decrease
of count rate with time in all of the apertures.  As in
Fig.~\ref{impact_lc}, the shape of the light curve is a function of
aperture.  Comparison with the $10.5~\mu$m light curves of
\citet{Sugita:2005} also suggests that the light curves are a function
of wavelength, as the infrared is mostly sensitive to larger, slower
moving grains.  In our case, we note that the total flux, as measured
in the 200 and 250 pixel radius apertures appears to remain constant
from about 1 to 5~hours after impact.  We use the count rates in these
apertures and the photometric calibration of \citet{Sirianni:2005} to
estimate the total $V$-band flux from the ejecta plume as $1.11 \times
10^{-14}$~\ergsa\ at 5888~\AA, which corresponds to a $V$ magnitude
of 13.7.
 
\placefigure{post_lc}

\subsection{Dust Content}

From the derived total flux we can estimate the effective surface area
of the ejected material, which together with an assumption about the
size distribution of the grains can provide a constraint on the total
mass of the solid ejecta.  Using the same geometric albedo (0.04) and
phase function (0.04 mag deg$^{-1}$) that was used to derive the
photometric size of the nucleus, the observed flux corresponds to an area of
960~km$^2$.  If instead we use the assumptions of \citet{Kuppers:2005},
a scattering albedo of 0.10 \citep[corresponding to a geometric albedo
of 0.025,][]{Hanner:1981}, and no phase function, we find an area of
340~km$^2$, in excellent agreement with the value of ($330 \pm
30$)~km$^2$ quoted by \citeauthor{Kuppers:2005}.  The semi-circular
area of the 200 pixel radius aperture is $1.65 \times 10^7$~km$^2$, so
the filling factor is $\leq 10^{-4}$ and optical depth effects can be
neglected.  We can also compare this result with the total flux
observed following the natural outburst of June 14 (Fig.~\ref{june})
which we find to be smaller by a factor of 7.1.  Correcting for the
heliocentric and geocentric distances on the two dates, the ratio of
ejected material is 8.6.  We can scale the impactor mass by this
factor, and find that a meteoroid of $\sim$40~kg mass, at the velocity
of Deep Impact, would be needed to produce such an outburst, quite
unlikely in view of the frequent nature and apparent periodicity of the
outbursts \citep{A'Hearn:2005b}, in agreement with the conclusion
reached by \citeauthor{Kuppers:2005}

While it is possible to use the derived area to calculate the total
ejecta volume (and then mass) by assuming a power law distribution of
particle sizes, visible imaging samples only a small portion of the
total mass distribution so that small uncertainties in the assumed size
distribution can lead to large differences in the final results.  This
accounts for the discrepancy between the mass derived by
\citeauthor{Kuppers:2005} and smaller values derived from infrared
imaging and spectroscopy \citep{Sugita:2005,Harker:2005}.  It is also
possible to use the rms expansion velocity derived above to
constrain the total mass using energy arguments
provided that we can determine what fraction of the kinetic
energy of the impactor is transferred to the solid ejecta during the
excavation of the crater.  However, we note the caveat that the velocity
distribution of Fig.~\ref{velocity} is only an approximate velocity
distribution because it is weighted by the velocity dependence of the
mass (and consequently, the area) of the scattering particles.  Thus,
detailed modeling, drawing upon additional imaging data in other
spectral bands, is needed to properly address this issue.

\section{CONCLUSION}

We have presented an overview of the visible imaging obtained with the
Advanced Camera for Surveys High Resolution Channel on the {\it Hubble
Space Telescope} before, during and following the encounter of the Deep
Impact spacecraft with comet 9P/Tempel~1.  In addition, we
serendipitously observed a natural outburst of the comet on 2005 June
14, whose characteristics may be compared with those of the ejecta
produced by the impact.  From the data we have obtained a measurement
of the photometric size of the nucleus, determined the ejecta expansion
velocity distribution, and followed the temporal evolution of the
morphology of the visible brightness of the ejecta cloud.  The high
spatial resolution and image quality of the HRC images should make
possible detailed kinematical modeling of the ejecta plume that may
serve to constrain the physical properties of the ejecta particles.

\begin{center}{\bf ACKNOWLEDGMENTS}\end{center}

This work is based on observations with the NASA/ESA {\it Hubble
Space Telescope} obtained at the Space Telescope Science Institute,
which is operated by the Association of Universities for Research in
Astronomy (AURA), Inc., under NASA contract NAS 5-26555.
We thank Ian Jordan, Ron Gilliland, and Charles Proffitt (STScI) for
the planning and successful execution of the \hst\ program; Eddie
Bergeron, Max Mutchler, Zolt Levay (STScI) and Ken Anderson (JHU) for
the rapid response and production of properly corrected images; Cheryl
Gundy, Lisa Frattare, Ray Villard, Mario Livio, and countless others at
STScI for the July~4 logistics; and STScI/JPL for coordination of
impact time before the end of \hst\ visibility.  This work was
supported by grant GO-10144.01-A from the Space Telescope Science
Institute.


\clearpage

\renewcommand\baselinestretch{1.1}%

\begin{table}
\begin{center}
\caption{Summary of ACS/HRC observations of comet 9P/Tempel~1. The
\hst\ visit number is the two digits preceding *** in the rootname.
The exposure sequence for each visit is described in the text.
Individual exposure IDs are given in the text where used.  The
parameters given correspond to the start time of the first exposure of
each visit.  \label{obs_table}}
\medskip
\begin{tabular}{@{}ccccccc@{}}
\tableline\tableline
Visit & Mode & Date & Start & $r$ & $\Delta$ & Phase \\
Rootname & &  & Time (UT) & (AU) & (AU) & Angle (\deg) \\
\tableline
\multicolumn{7}{c}{June pre-impact} \\
J8Z301***  & HRC F606W + PR200L	& 2005-06-13  &	05:36:11 & 1.522 & 0.799 & 37.7 \\
J8Z302***  & HRC F606W + HRC-512 & 2005-06-13  & 15:11:58 & 1.522 & 0.800 & 37.8 \\  
J8Z303***  & HRC F606W + PR200L & 2005-06-14  &	07:11:36 & 1.521 & 0.803 & 37.9 \\
J8Z304***  & HRC F606W + HRC-512 & 2005-06-14  & 14:09:20 & 1.521 & 0.804 & 38.0 \\
\tableline
\multicolumn{7}{c}{Impact epoch} \\
J9A801*** & HRC F606W + PR200L  & 2005-07-02  &	00:32:22 & 1.507 & 0.883 & 40.7 \\
J9A803*** & HRC F606W + PR200L  & 2005-07-02  &	22:55:32 & 1.506 & 0.887 & 40.8 \\
J9A805*** & HRC-512 F606W (impact)    & 2005-07-04  & 05:19:10 & 1.506 & 0.894 & 40.9 \\
J9A806*** & HRC F606W + PR200L  & 2005-07-04  &	06:54:26 & 1.506 & 0.894 & 40.9 \\
J9A808*** & HRC F606W + PR200L  & 2005-07-04  &	10:06:20 & 1.506 & 0.895 & 41.0 \\
J9A810*** & HRC F606W + PR200L  & 2005-07-05  &	00:29:54 & 1.506 & 0.898 & 41.0 \\
J9A812*** & HRC F606W + PR200L  & 2005-07-11  &	18:03:11 & 1.508 & 0.935 & 41.4 \\
J9A822*** & HRC F606W + PR200L  & 2005-07-16  &	22:55:25 & 1.511 & 0.966 & 41.6 \\
\tableline
\tableline
\end{tabular}
\end{center}
\end{table}

\clearpage
\renewcommand\baselinestretch{1.2}%

\clearpage 
\begin{center}{\bf FIGURE CAPTIONS}\end{center}

\figcaption{Left: Mean of four exposures from visit 01, J9A801DZQ,
J9A801E0Q, J9A801E5Q, and J9A801E6Q, each 140~s, taken about 54 hours
pre-impact.  Right, Mean of exposures J9A806EQQ and J9A806ERQ, both
140~s, taken approximately one hour after impact.  The legend in the
right panel gives the start time of the first exposure.  \label{full}}

\figcaption{Azimuthally averaged radial profile of an image taken
immediately prior to impact (J9A805EAQ).  The green line is a dust coma
model (proportional to $\rho^{-1}$) convolved with the instrumental
point-spread function while the red line is the difference between
the data and the coma model, which yields an estimate for the
contribution from the nucleus.  The pixel size is $0.\!''025$. 
\label{nucleus}}

\figcaption{Images of the natural outburst on 2005 June 14 created by
taking the ratio of exposures from the fourth visit to a mean of two
exposures obtained $\sim$7~hours earlier (J8Z303G2Q and J8Z303G3Q).  The
two images were taken 67 minutes apart.  Left: Mean of J8Z304I7Q and
J8Z304I8Q.  Right:  Mean of J8Z304IBQ and J8Z304ICQ.  All exposures
were 300~s.  The exposure start times are indicated in the figure.  The
``tongue depresser'' feature in the NW quadrant is due to an occulting
finger in the HRC.  \label{june}}

\figcaption{Contour plots of the two ratio images shown in
Fig.~\ref{june}.  The fan appears symmetric about a position angle of
311\deg, while the position angle of the Sun is 295\deg.  Semi-circles,
centered on the nucleus are shown, of radius 2330~km and 2910~km,
respectively, for the left and right plots.  \label{contour_june}}

\figcaption{Ratio images of the 40-s exposures taken during and in the
13.6~min immediately following impact to the mean of the last four prior
exposures.  The first exposure, J9A805EBQ began 23~s before the time of
impact as seen from Earth.  The exposure mid-points are 75~s apart.  \label{visit05}}

\figcaption{Contour plots of four of the ratio images shown in
Fig.~\ref{visit05}.  \label{contour05}}

\figcaption{Images created by taking the ratio of post-impact exposures
to the pre-impact image shown in the left panel of Fig.~\ref{full}.
Left:  Mean of J9A806EQQ and J9A806ERQ, each 140~s.  Right: Mean of
J9A806EWQ and J9A806EXQ, each 140~s.  The exposure start times are
indicated in the figure.  The exposure mid-points are 68.2 (left) and
92.2~min (right) after impact.  \label{visit06}}

\figcaption{Same as Fig.~\ref{visit06}.  Left:  Mean of J9A808F9Q and
J9A808FAQ, each 140~s.  Right: Mean of J9A808FFQ and J9A808FGQ, each
140~s.  The exposure mid-points are 260.1 (left) and 284.1~min (right)
after impact. \label{visit08}}

\figcaption{Contour plots of the two ratio images shown in
Fig.~\ref{visit06}.  The fan appears symmetric about a position angle
of 240\deg.  Semi-circles, centered on the nucleus are shown, of radius
1210~km and 1610~km, respectively, for the left and right plots.
\label{contour06}}

\figcaption{Radial profiles of the pre- and post-impact images used in the
ratio contour plots shown in Fig.~\ref{contour06}.  The range of
position angle is 180 to 300\deg.  Black: pre-impact; Red: post-impact;
Blue: difference between post- and pre-impact, multiplied by radial
distance (in pixel units).  The dashed curve gives the fraction of
total flux (multiplied by 1000) remaining outside a given radial
distance.\label{rad06}}

\figcaption{Normalized velocity distributions derived from the blue curves in 
Fig.~\ref{rad06} (Black: left panel; Red: Right panel).  \label{velocity}}

\figcaption{Azimuthal profiles of the pre-impact (red, $\times 10$) and
post-impact (black) images used in the ratio contour plot shown in the
left panel of Fig.~\ref{contour06}.  The range of radial distance is
200 to 650 km.  \label{az06}}

\figcaption{Light curves based on aperture photometry.  Left: impact
orbit.  Right: the following \hst\ orbit.  The pre-impact count rates
are 1.76, 2.78, 3.95, and 6.96, all $\times 10^3$~counts~s$^{-1}$, for
the 40, 80, 160, and 400~km radius apertures, respectively, and have
been subtracted from the data points.  \label{impact_lc}}

\figcaption{Light curves based on aperture photometry through 20~h
following impact.  The pre-impact values have not been subtracted as in
Fig.~\ref{impact_lc}.  \label{post_lc}}

\setcounter{figure}{0}
\clearpage

\begin{figure*}
\begin{center}
\epsscale{1.0}
\rotatebox{0.}{
\plotone{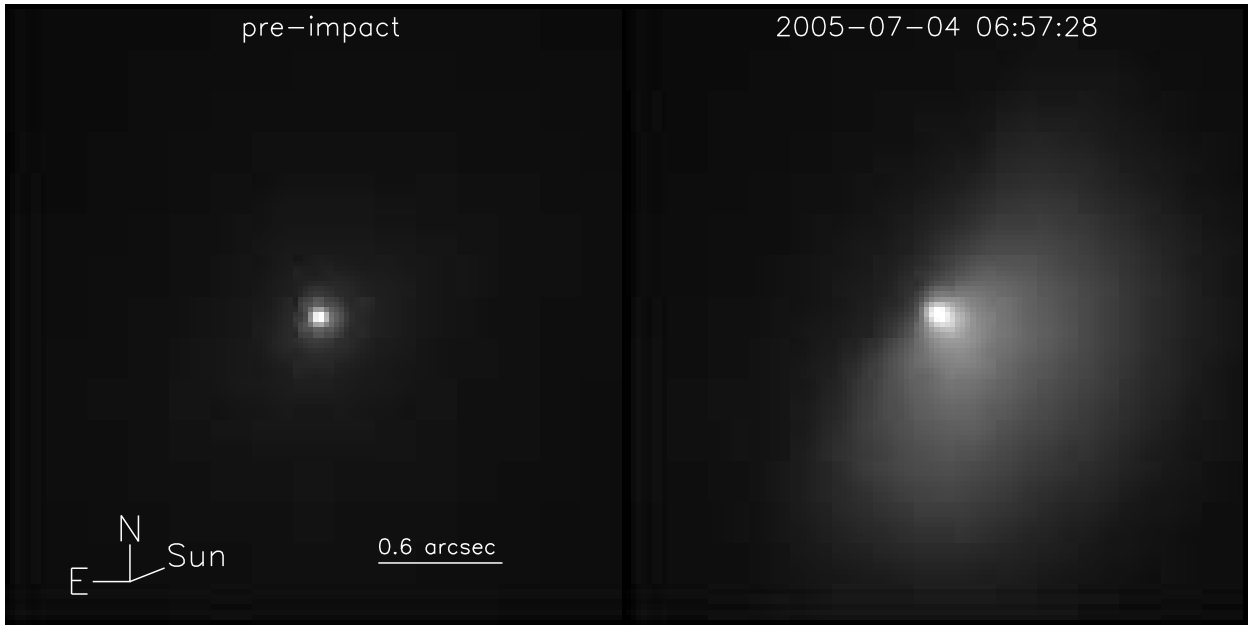}}
\caption{Left: Mean of four exposures from visit 01, J9A801DZQ,
J9A801E0Q, J9A801E5Q, and J9A801E6Q, each 140~s, taken about 54 hours
pre-impact.  Right, Mean of exposures J9A806EQQ and J9A806ERQ, both
140~s, taken approximately one hour after impact.  The legend in the
right panel gives the start time of the first exposure.  }
\end{center}
\end{figure*}

\begin{figure*}
\begin{center}
\epsscale{1.0}
\rotatebox{0.}{
\plotone{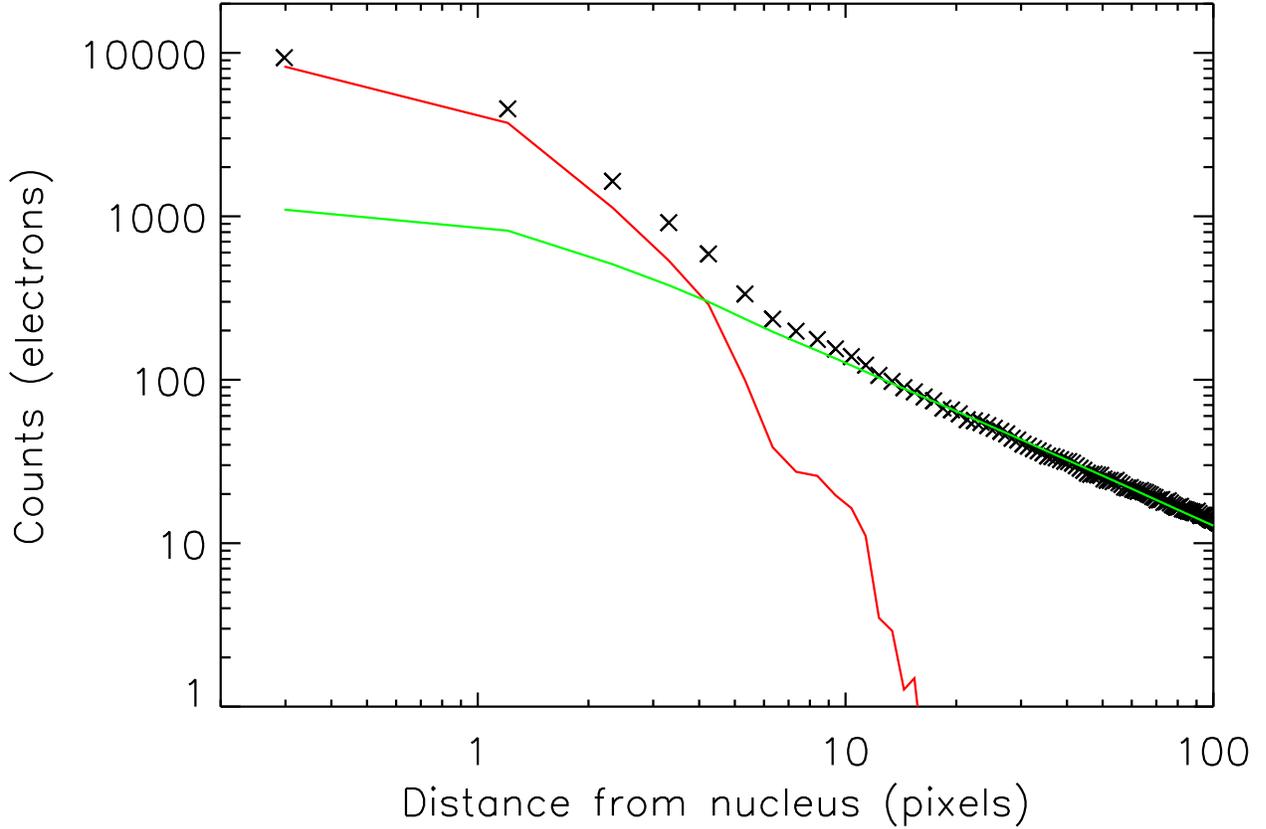}}
\caption{Azimuthally averaged radial profile of an image taken
immediately prior to impact (J9A805EAQ).  The green line is a dust coma
model (proportional to $\rho^{-1}$) convolved with the instrumental
point-spread function while the red line is the difference between
the data and the coma model, which yields an estimate for the
contribution from the nucleus.  The pixel size is $0.\!''025$.} 
\end{center}
\end{figure*}

\begin{figure*}
\begin{center}
\epsscale{0.95}
\rotatebox{0.}{
\plotone{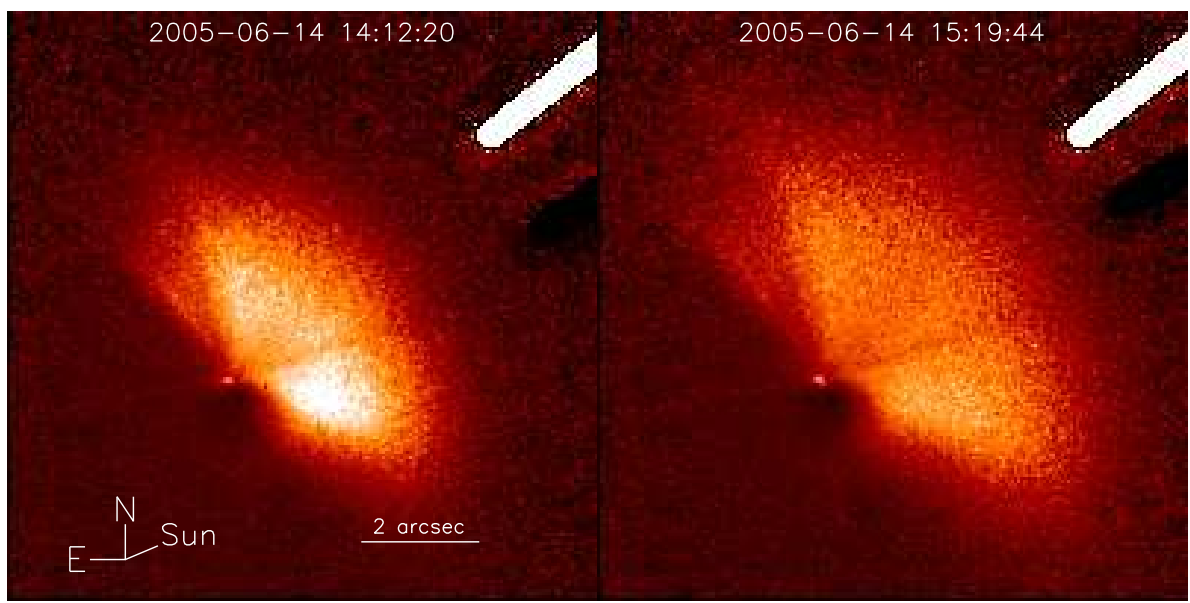}}
\vspace{0.1in}
\caption{Images of the natural outburst on 2005 June 14 created by
taking the ratio of exposures from the fourth visit to a mean of two
exposures obtained $\sim$7~hours earlier (J8Z303G2Q and J8Z303G3Q).  The
two images were taken 67 minutes apart.  Left: Mean of J8Z304I7Q and
J8Z304I8Q.  Right:  Mean of J8Z304IBQ and J8Z304ICQ.  All exposures
were 300~s.  The exposure start times are indicated in the figure.  The
``tongue depresser'' feature in the NW quadrant is due to an occulting
finger in the HRC.}
\end{center}
\end{figure*}

\begin{figure*}
\begin{center}
\epsscale{1.20}
\rotatebox{0}{
\plottwo{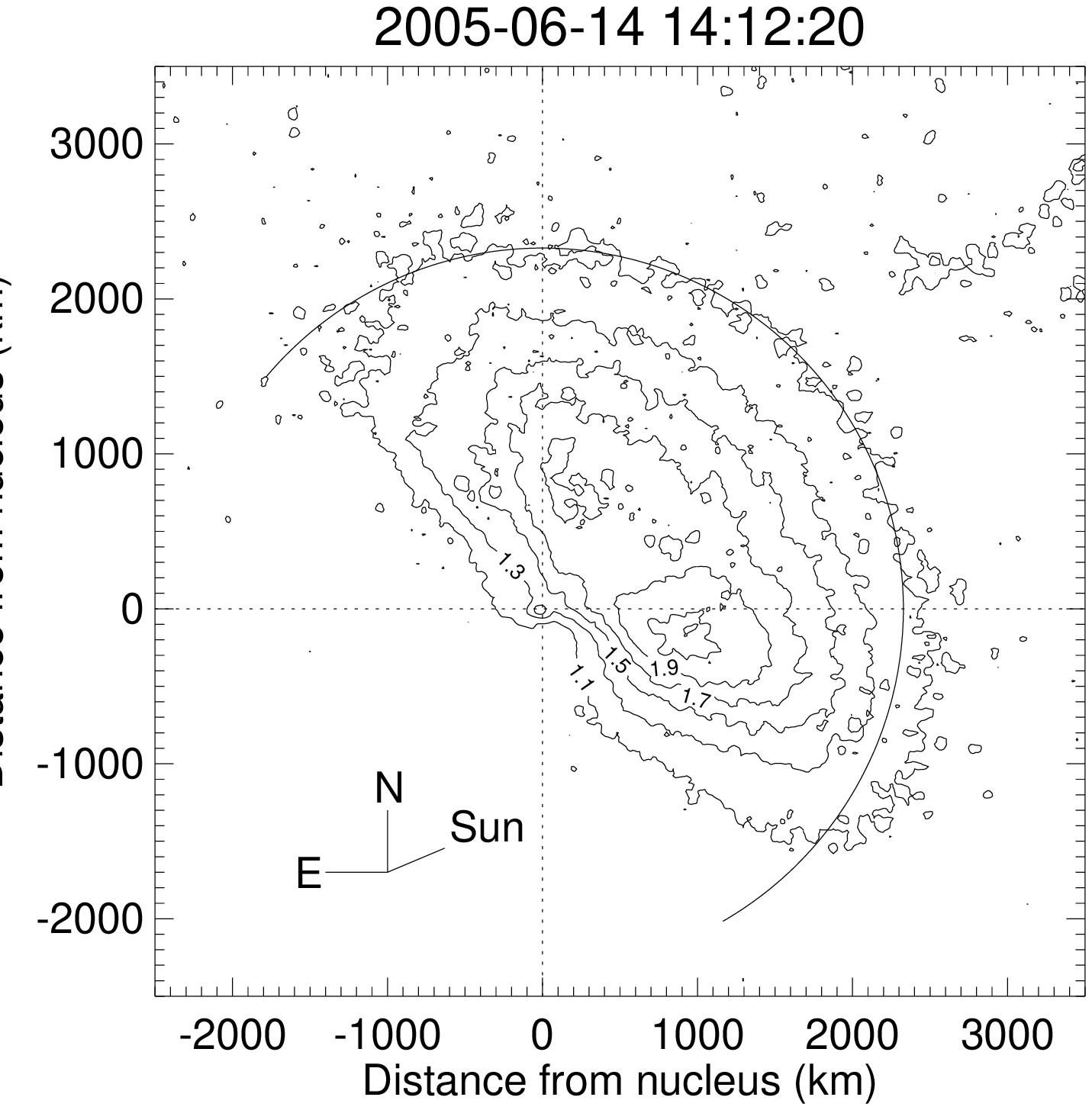}{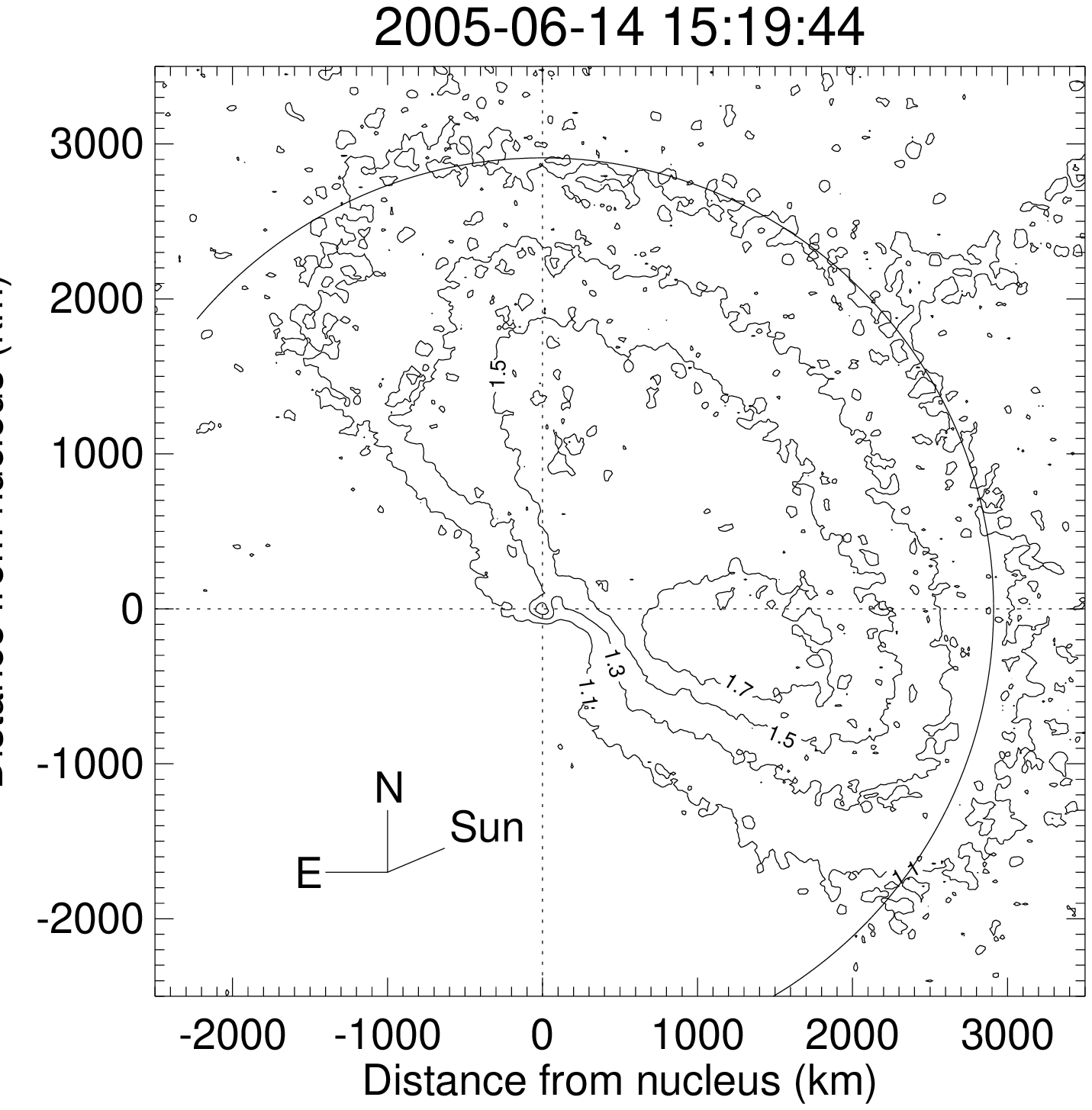}}
\caption{Contour plots of the two ratio images shown in
Fig.~\ref{june}.  The fan appears symmetric about a position angle of
311\deg, while the position angle of the Sun is 295\deg.  Semi-circles,
centered on the nucleus are shown, of radius 2330~km and 2910~km,
respectively, for the left and right plots.}
\end{center}
\end{figure*}

\begin{figure*}
\begin{center}
\epsscale{0.75}
\rotatebox{90}{
\plotone{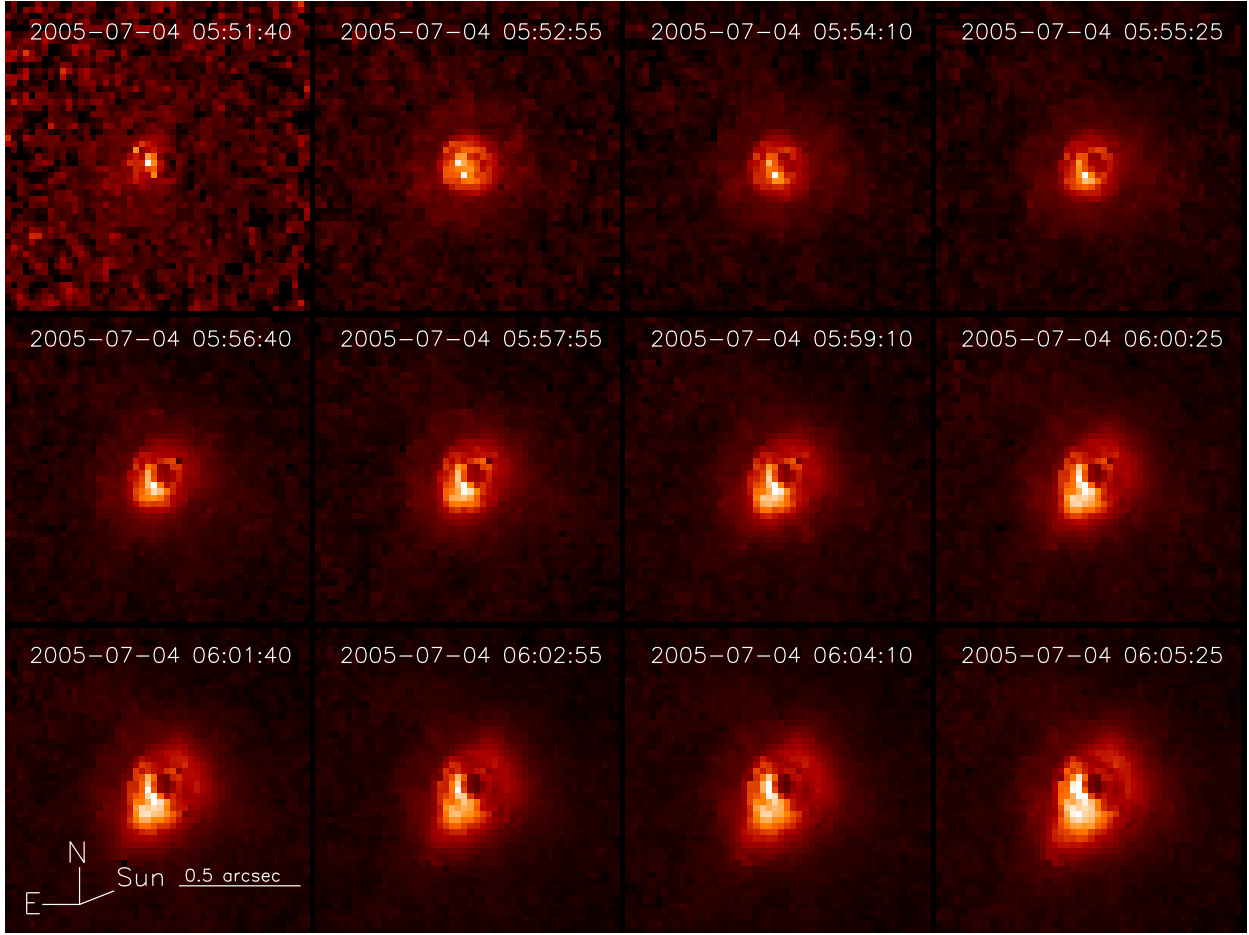}}
\vspace{0.1in}
\caption{Ratio images of the 40-s exposures taken during and in the
13.6~min immediately following impact to the mean of the last four prior
exposures.  The first exposure, J9A805EBQ began 23~s before the time of
impact as seen from Earth.  The exposure mid-points are 75~s apart.}
\end{center}
\end{figure*}

\begin{figure*}
\begin{center}
\epsscale{1.0}
\rotatebox{0}{
\plotone{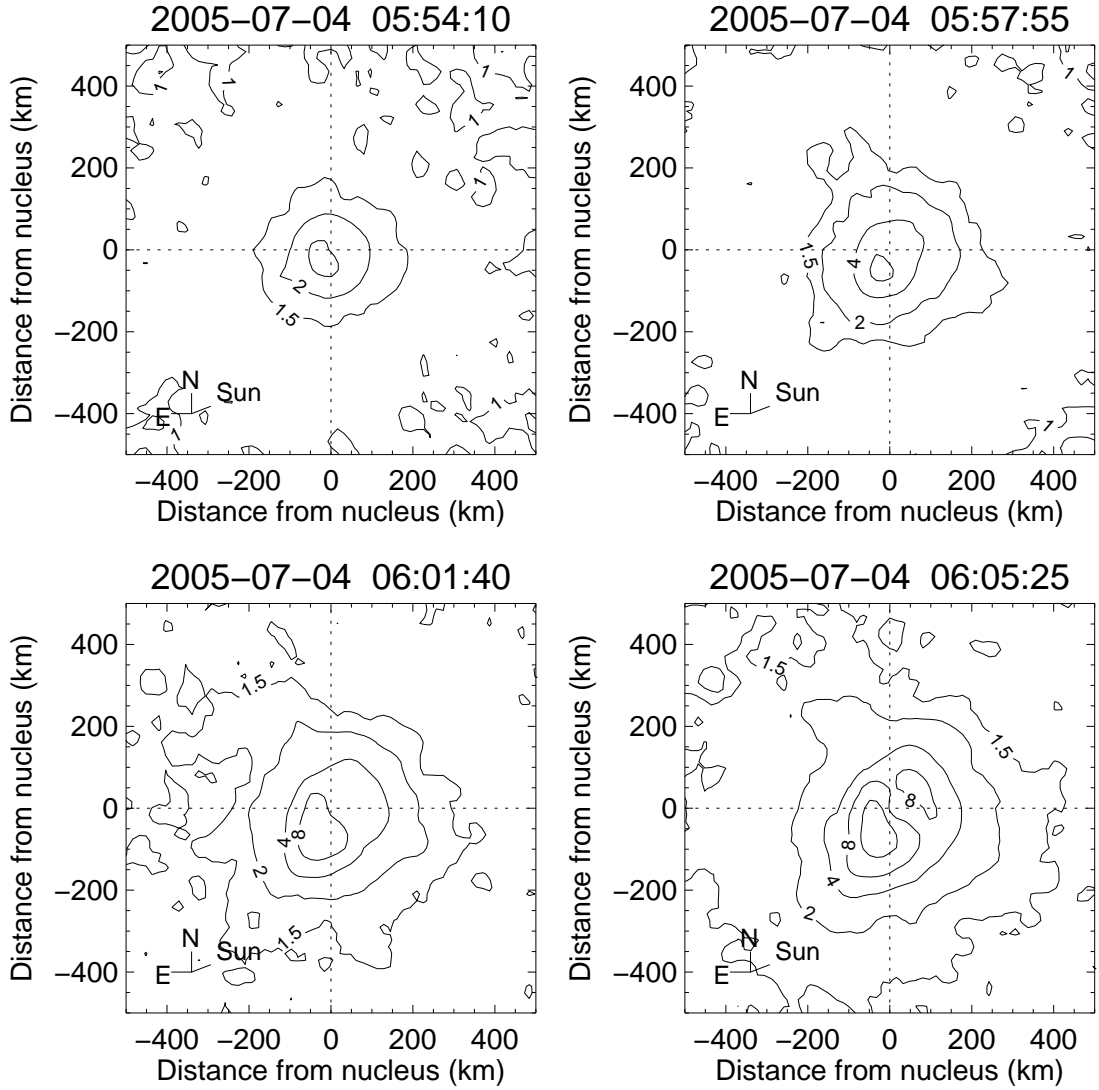}}
\vspace{-0.5in}
\caption{Contour plots of four of the ratio images shown in
Fig.~\ref{visit05}.}
\end{center}
\end{figure*}

\begin{figure*}
\begin{center}
\epsscale{1.0}
\rotatebox{0.}{
\plotone{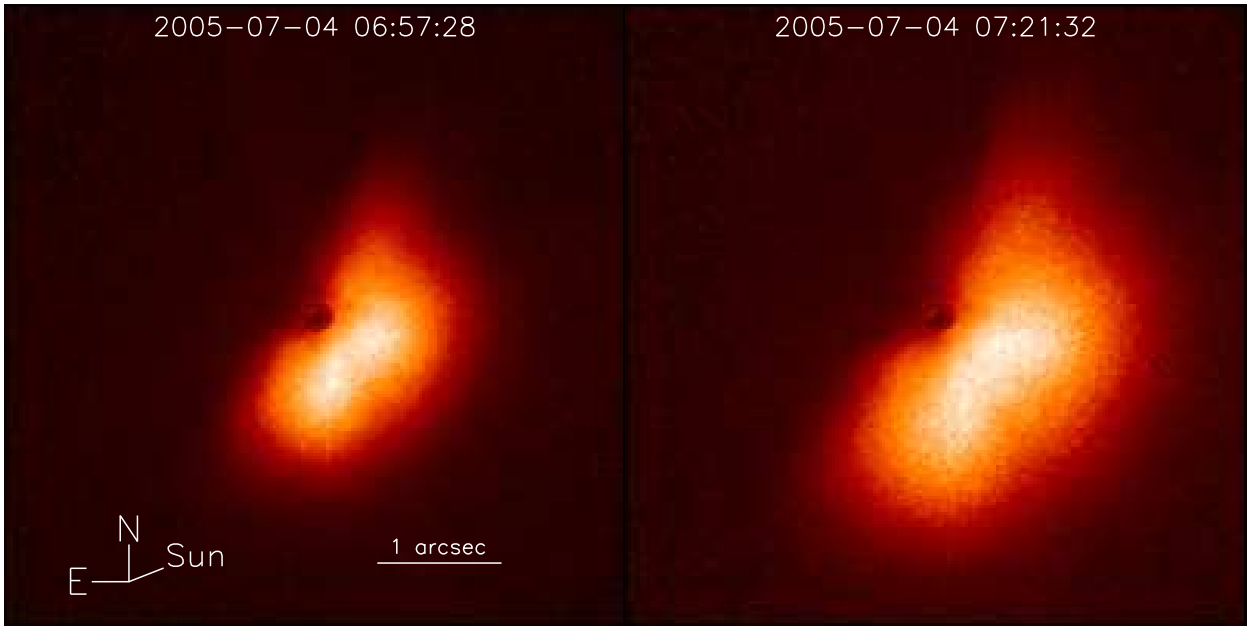}}
\caption{Images created by taking the ratio of post-impact exposures to
the pre-impact image shown in the left panel of Fig.~\ref{full}.  Left:
Mean of J9A806EQQ and J9A806ERQ, each 140~s.  Right: Mean of J9A806EWQ
and J9A806EXQ, each 140~s.  The exposure start times are indicated in
the figure.  The exposure mid-points are 68.2 (left) and 92.2~min
(right) after impact.}
\end{center}
\end{figure*}

\begin{figure*}
\begin{center}
\epsscale{1.0}
\rotatebox{0.}{
\plotone{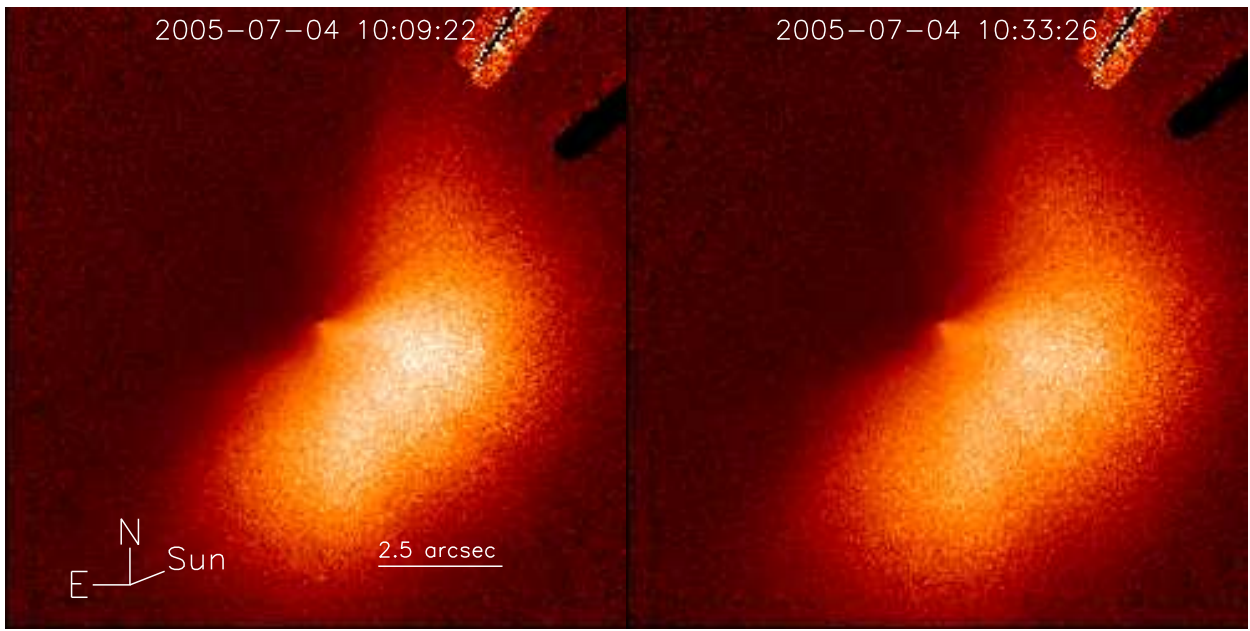}}
\caption{Same as Fig.~\ref{visit06}.  Left:  Mean of J9A808F9Q and
J9A808FAQ, each 140~s.  Right: Mean of J9A808FFQ and J9A808FGQ, each
140~s.  The exposure mid-points are 260.1 (left) and 284.1~min (right)
after impact.}
\end{center}
\end{figure*}

\begin{figure*}
\begin{center}
\epsscale{1.20}
\rotatebox{0}{
\plottwo{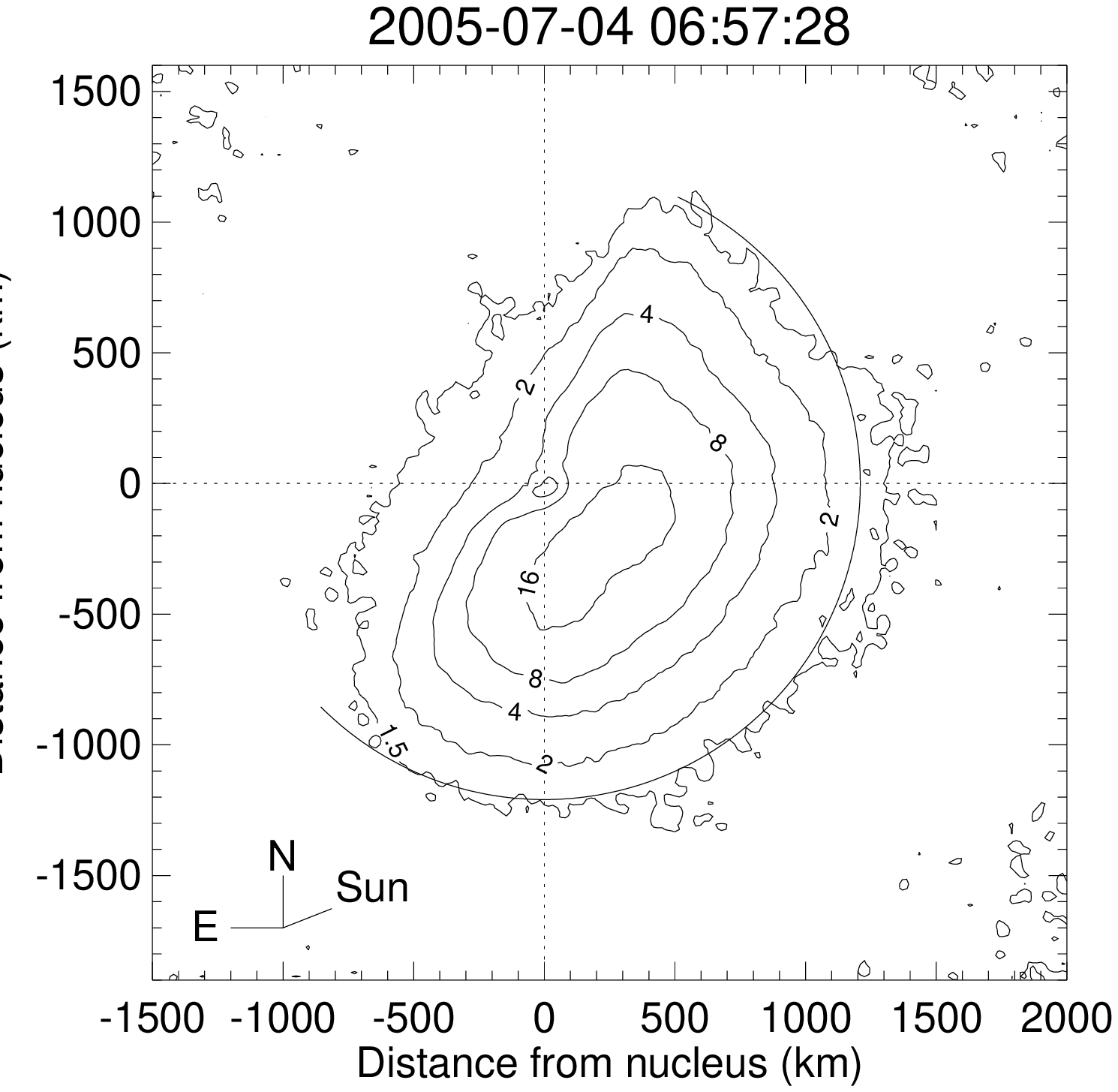}{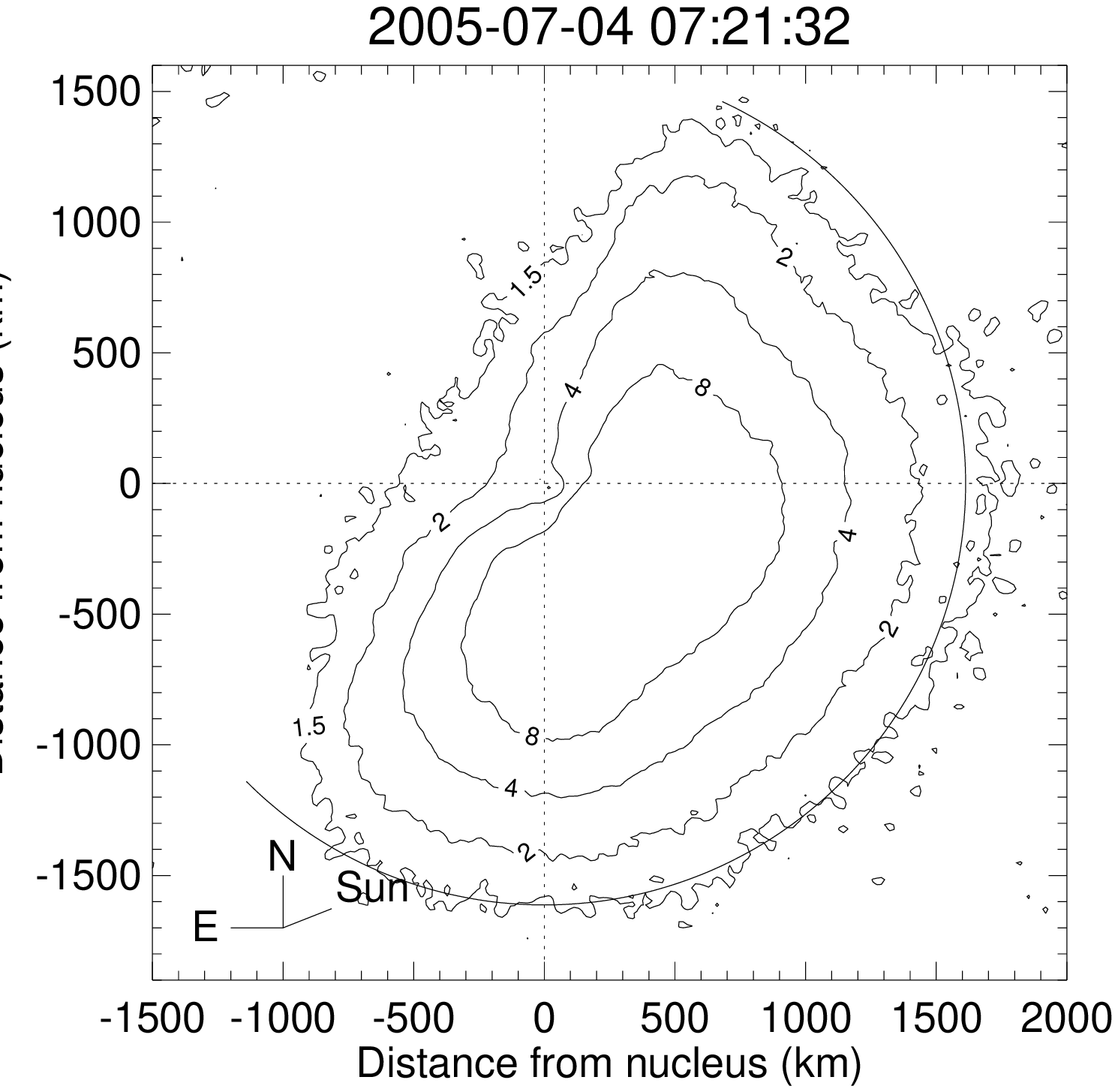}}
\caption{Contour plots of the two ratio images shown in
Fig.~\ref{visit06}.  The fan appears symmetric about a position angle
of 240\deg.  Semi-circles, centered on the nucleus are shown, of radius
1210~km and 1610~km, respectively, for the left and right plots. }
\end{center}
\end{figure*}

\begin{figure*}
\begin{center}
\epsscale{1.10}
\rotatebox{0}{
\plottwo{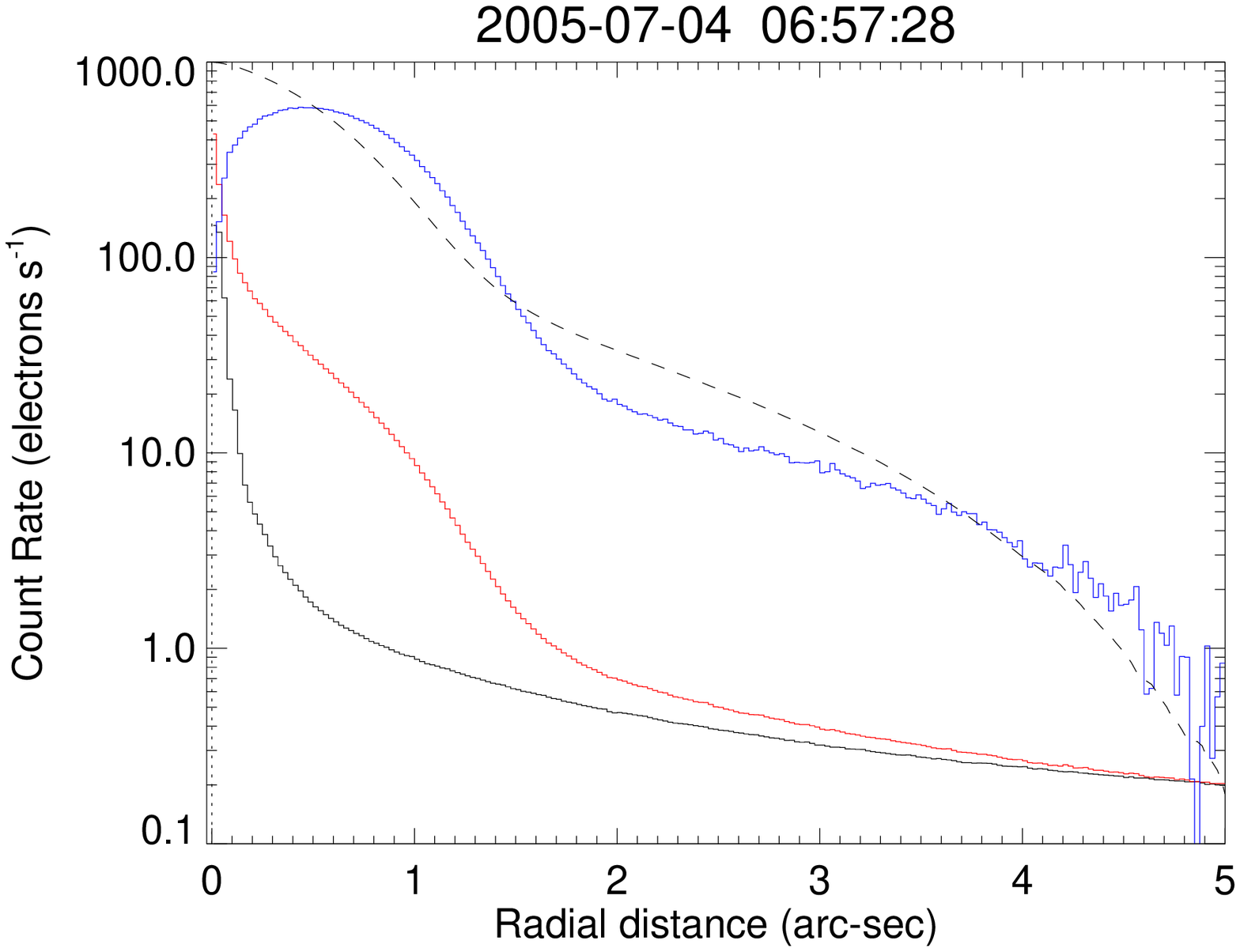}{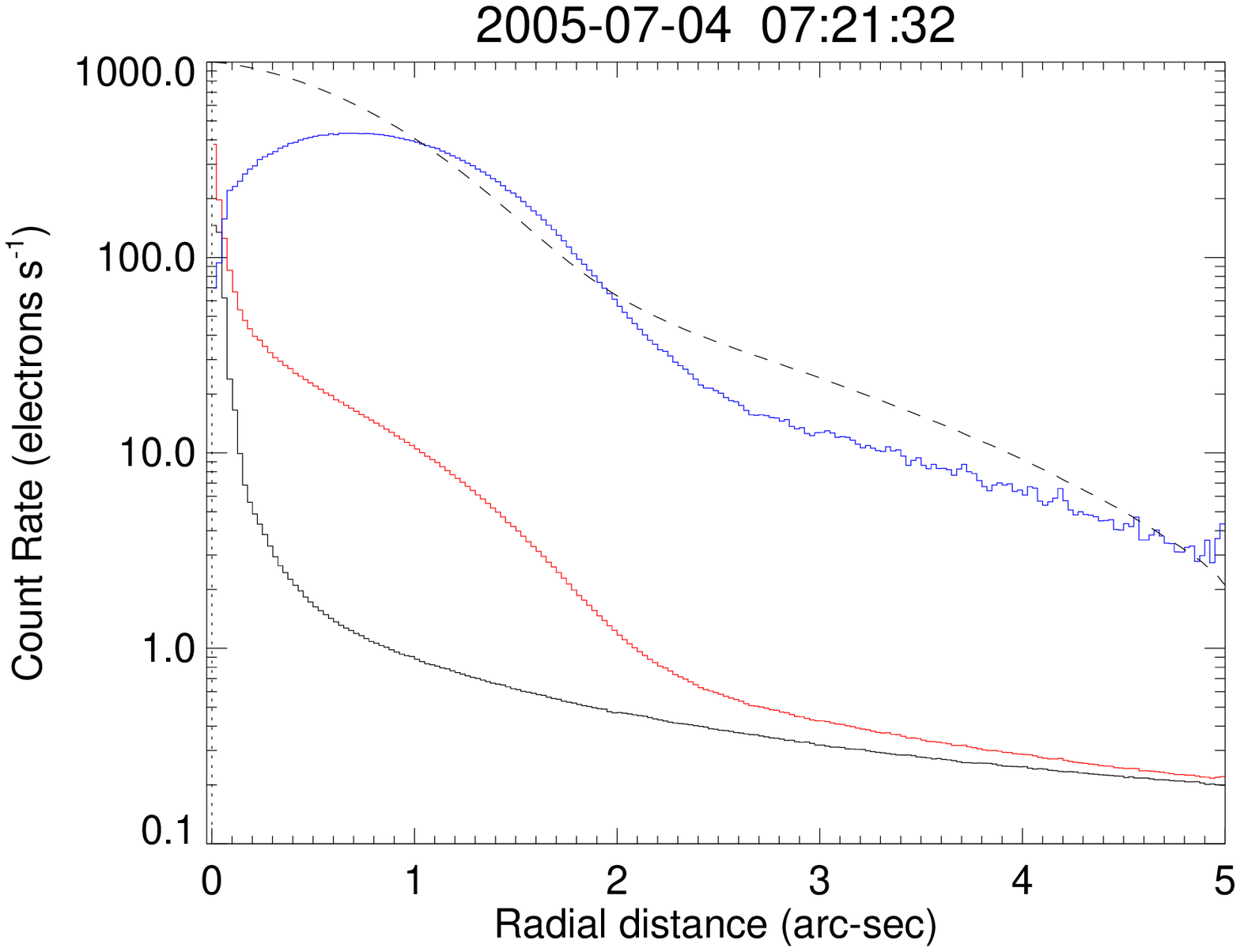}}
\caption{Radial profiles of the pre- and post-impact images used in the
ratio contour plots shown in Fig.~\ref{contour06}.  The range of
position angle is 180 to 300\deg.  Black: pre-impact; Red: post-impact;
Blue: difference between post- and pre-impact, multiplied by radial
distance (in pixel units).  The dashed curve gives the fraction of
total flux (multiplied by 1000) remaining outside a given radial
distance. }
\end{center}
\end{figure*}

\begin{figure*}
\begin{center}
\epsscale{1.0}
\rotatebox{0}{
\plotone{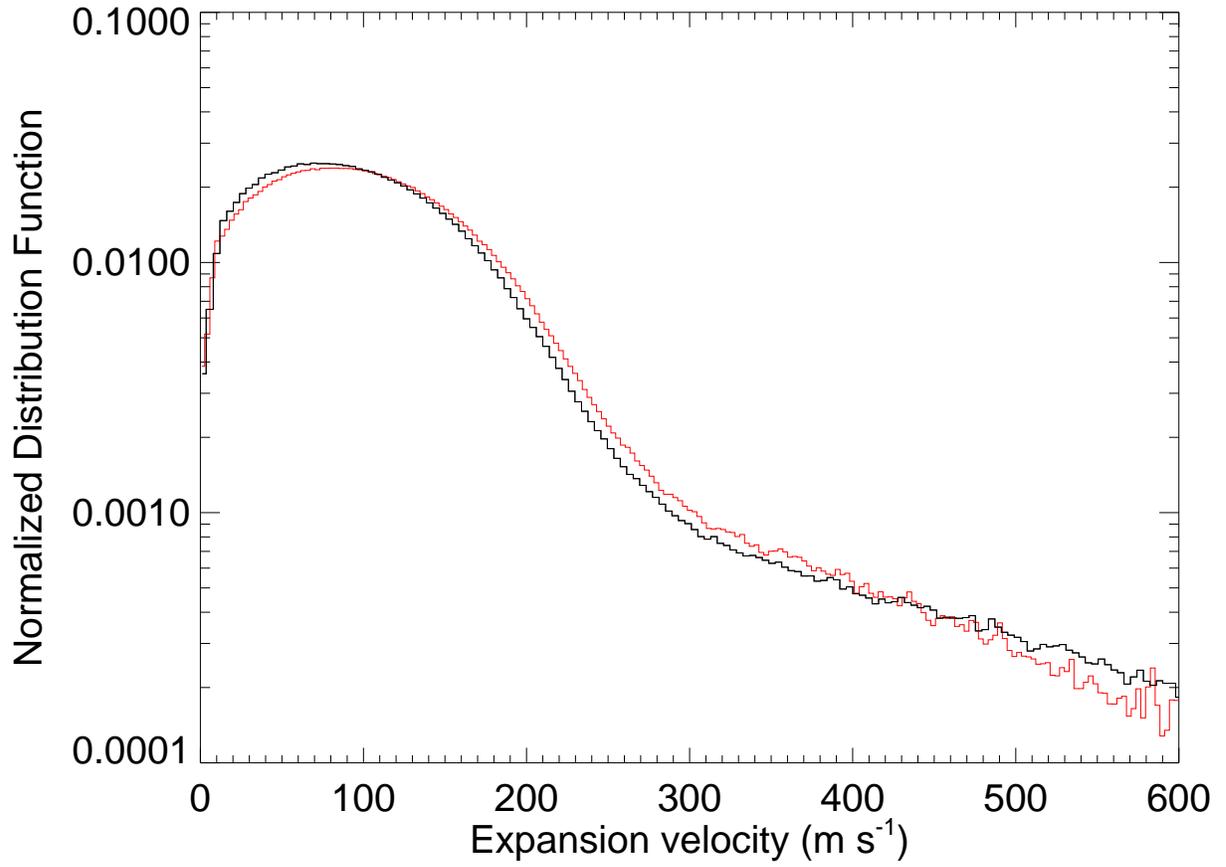}}
\caption{Normalized velocity distributions derived from the blue curves in 
Fig.~\ref{rad06} (Black: left panel; Red: Right panel). }
\end{center}
\end{figure*}

\begin{figure*}
\begin{center}
\epsscale{1.0}
\rotatebox{0}{
\plotone{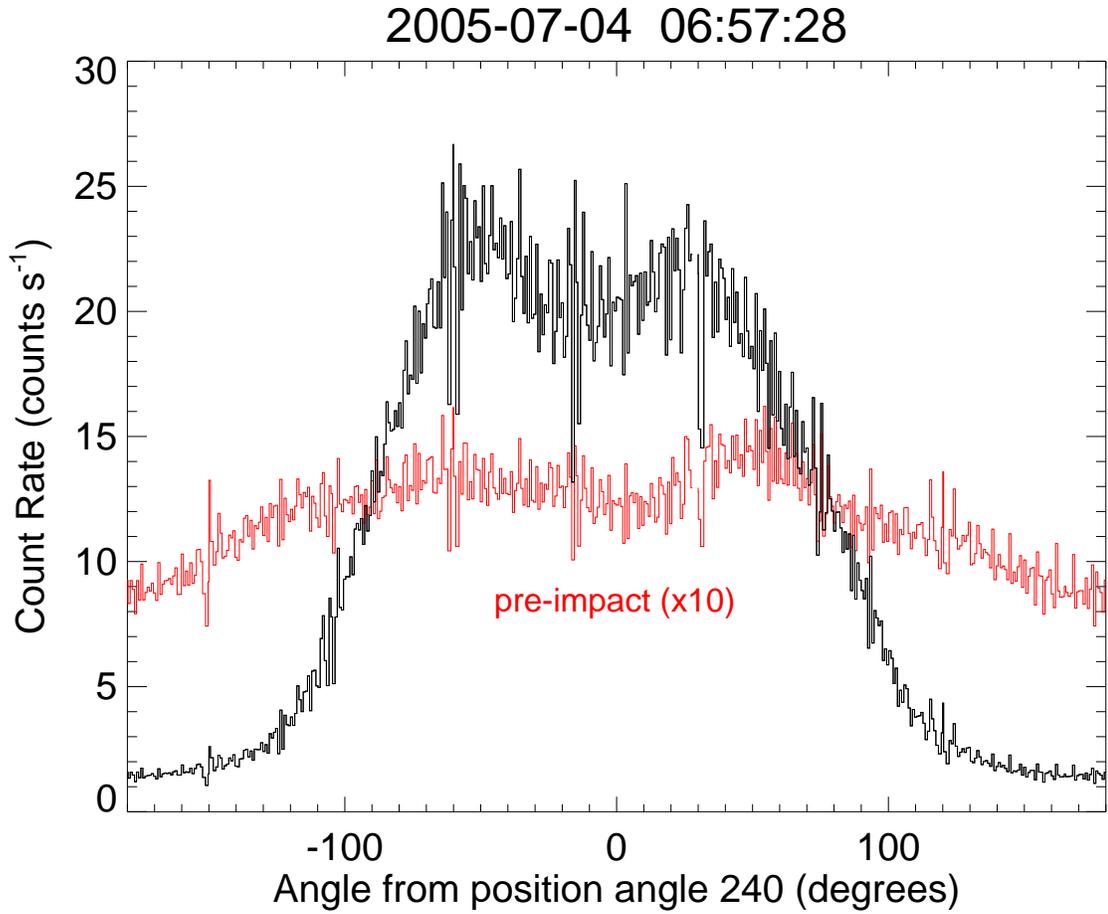}}
\caption{Azimuthal profiles of the pre-impact (red, $\times 10$) and
post-impact (black) images used in the ratio contour plot shown in the
left panel of Fig.~\ref{contour06}.  The range of radial distance is
200 to 650 km.}
\end{center}
\end{figure*}

\begin{figure*}
\begin{center}
\epsscale{1.0}
\rotatebox{0.}{
\plotone{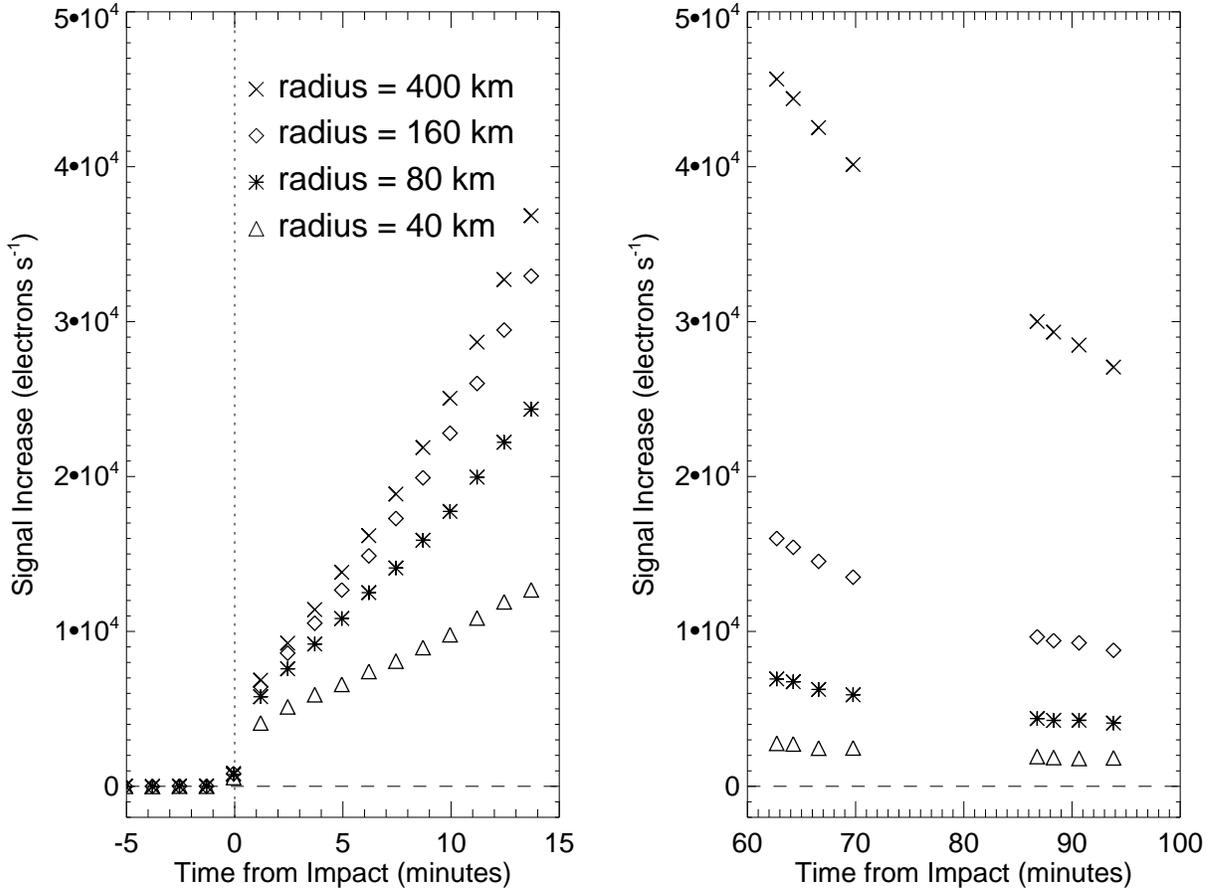}}
\caption{Light curves based on aperture photometry.  Left: impact
orbit.  Right: the following \hst\ orbit.  The pre-impact count rates
are 1.76, 2.78, 3.95, and 6.96, all $\times 10^3$~counts~s$^{-1}$, for
the 40, 80, 160, and 400~km radius apertures, respectively, and have
been subtracted from the data points. }
\end{center}
\end{figure*}

\begin{figure*}
\begin{center}
\epsscale{1.0}
\rotatebox{0.}{
\plotone{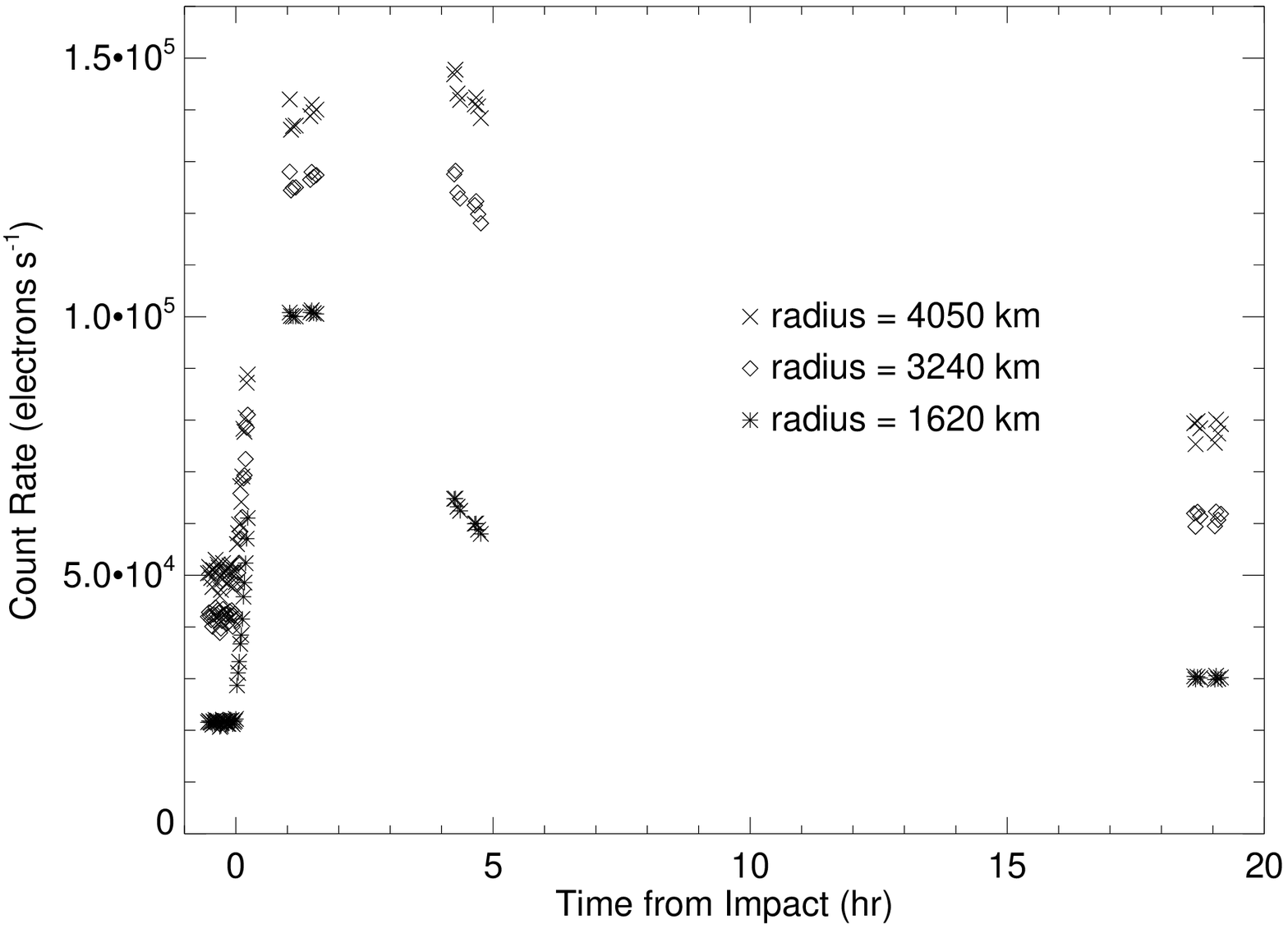}}
\caption{Light curves based on aperture photometry through 20~h
following impact.  The pre-impact values have not been subtracted as in
Fig.~\ref{impact_lc}. }
\end{center}
\end{figure*}


\begin{thebibliography}{}

\bibitem[A'Hearn et al.(2005a)]{A'Hearn:2005a} A'Hearn, M.~F., Belton, 
M.~J.~S., Delamere, A., Blume, W.~H.\ 2005a.\ Deep Impact: A Large-Scale 
Active Experiment on a Cometary Nucleus.\ Space Science Reviews 117, 1--21. 
 
\bibitem[A'Hearn et al.(2005b)]{A'Hearn:2005b} A'Hearn, M.~F., and 32 
colleagues 2005b.\ Deep Impact: Excavating Comet Tempel 1.\ Science 310, 
258--264. 
 
\bibitem[Carcich and Elliott(2006)]{Carcich:2006} Carcich, B., Elliott, 
G.\ 2006.\ View of Impact Site from Earth.\ Deep Impact Team
Document. 
 
\bibitem[Feldman et al.(2006)]{Feldman:2006} Feldman, P.~D., Lupu,
R.~E., McCandliss, S.~R., Weaver, H.~A., A'Hearn, M.~F., Belton,
M.~J.~S., Meech, K.~J.\ 2006.\ Carbon Monoxide in Comet 9P/Tempel~1
before and after the Deep Impact Encounter.\ Astrophysical
Journal (Letters) 647, L61.
 
\bibitem[Hanner et al.(1981)]{Hanner:1981} Hanner, M.~S., Giese, 
R.~H., Weiss, K., Zerull, R.\ 1981.\ On the definition of albedo and 
application to irregular particles.\ Astronomy and Astrophysics 104, 42-46. 

\bibitem[Harker et al.(2005)]{Harker:2005} Harker, D.~E., Woodward, 
C.~E., Wooden, D.~H.\ 2005.\ The Dust Grains from 9P/Tempel 1 Before and 
After the Encounter with Deep Impact.\ Science 310, 278--280. 
 
\bibitem[Keller et al.(2005)]{Keller:2005} Keller, H.~U., and 11 
colleagues 2005.\ Deep Impact Observations by OSIRIS Onboard the Rosetta 
Spacecraft.\ Science 310, 281--283. 

\bibitem[K{\"u}ppers et al.(2005)]{Kuppers:2005} K{\"u}ppers, M., 
and 40 colleagues 2005.\ A large dust/ice ratio in the nucleus of comet 
9P/Tempel 1.\ Nature 437, 987--990. 
 
\bibitem[Lamy et al.(2001)]{Lamy:2001} Lamy, P.~L., Toth, I., 
A'Hearn, M.~F., Weaver, H.~A., Weissman, P.~R.\ 2001.\ Hubble Space 
Telescope Observations of the Nucleus of Comet 9P/Tempel 1.\ Icarus 154, 
337--344. 

\bibitem[Lamy et al.(2004)]{Lamy:2004} Lamy, P.~L., Toth, I., 
Fernandez, Y.~R., Weaver, H.~A.\ 2004.\ The sizes, shapes, albedos, and
colors of cometary nuclei, in Comets II, ed. M.~C. {Festou}, H.~A.
{Weaver}, \& H.~U. {Keller} (Tucson: Univ. of Arizona), 223--264.

\bibitem[Lara et al.(2006)]{Lara:2006} Lara, L.~M., Boehnhardt, 
H., Gredel, R., Guti{\'e}rrez, P.~J., Ortiz, J.~L., Rodrigo, R., 
Vidal-Nu{\~n}ez, M.~J.\ 2006.\ Pre-impact monitoring of Comet 9P/Tempel 1, 
the Deep Impact target.\ Astronomy and Astrophysics 445, 1151--1157. 

\bibitem[Meech et al.(2005)]{Meech:2005} Meech, K.~J., and 208 
colleagues 2005.\ Deep Impact: Observations from a Worldwide Earth-Based 
Campaign.\ Science 310, 265--269. 
 
\bibitem[Schleicher et al.(2006)]{Schleicher:2006} Schleicher, D.~G., 
Barnes, K.~L., Baugh, N.~F.\ 2006.\ Photometry and Imaging Results for 
Comet 9P/Tempel 1 and Deep Impact: Gas Production Rates, Postimpact Light 
Curves, and Ejecta Plume Morphology.\ Astronomical Journal 131, 1130--1137. 
 
\bibitem[Schultz et al.(2005)]{Schultz:2005} Schultz, P.~H., Ernst, 
C.~M., Anderson, J.~L.~B.\ 2005.\ Expectations for Crater Size and 
Photometric Evolution from the Deep Impact Collision.\ Space Science 
Reviews 117, 207--239. 
 
\bibitem[Sirianni et al.(2005)]{Sirianni:2005} Sirianni, M., and 13 
colleagues 2005.\ The Photometric Performance and Calibration of the Hubble 
Space Telescope Advanced Camera for Surveys.\ Publications of the 
Astronomical Society of the Pacific 117, 1049--1112. 

\bibitem[Sugita et al.(2005)]{Sugita:2005} Sugita, S., and 22 
colleagues 2005.\ Subaru Telescope Observations of Deep Impact.\ Science 
310, 274--278. 

\end{thebibliography}
\end{document}